\documentstyle[prl,aps,epsfig,multicol]{revtex}
\newcommand{\dpar}{D}
\newcommand{\dperp}{d}
\begin{document}
\def\posvar{{\bf r}}
\def\posexpo{\Theta}
\def\timexpo{x}
\def\phinn{\Omega}

\newif\ifmulticol
\multicoltrue
\def\mcbegin{\ifmulticol\begin{multicols}{2}\else\narrowtext\fi}
\def\mcend{\ifmulticol\end{multicols}\else\widetext\fi}

\draft
\title{Delocalization transitions of semi-flexible manifolds}
\author{Ralf Bundschuh${}^{1}$ and Michael L{\"a}ssig}
\address{
Max-Planck-Institut f\"ur Kolloid- und Grenzfl\"achenforschung,
Kantstr.~55, 14513 Teltow, Germany\\
${}^1$Current address: UCSD, Physics Department, 9500 Gilman Drive, La Jolla,
CA 92093-0319, USA}
\date{February 16, 1999}
\maketitle

\begin{abstract}
Semi-flexible manifolds such as fluid membranes or semi-flexible
polymers undergo delocalization transitions if they are subject to
attractive interactions.  We study manifolds with short-ranged
interactions by field-theoretic methods based on the operator product
expansion of local interaction fields. We apply this approach to manifolds in
a random potential. Randomness is always relevant for fluid membranes,
while for semi-flexible polymers there is a first order transition to
the strong coupling regime at a finite temperature.
\end{abstract}
\pacs{PACS number(s): 87.16.Dg, 64.60.Fr, 61.41.+e}

\mcbegin

Low dimensional manifolds play an important role in a variety of
different contexts, e.g., as soft matter objects or as domain
boundaries in condensed matter systems. They can perform large shape
fluctuations driven by entropy~\cite{lipo89b}. According to their
fluctuations they can be divided into two classes. {\em Flexible}
manifolds, such as interfaces, polymerized membranes, and long
polymers, fluctuate under a {\em tension} controlling their area or
length.  The other class is governed by {\em bending energy}; i.e.,
regions of high {\em curvature} are penalized.  Examples are polymers
not much longer than their persistence length, like actin or
DNA, and fluid membranes. These objects are stiffer, and we call
them {\em semi-flexible\/} manifolds.

Whenever a fluctuating manifold is attracted towards some
other ``defect'' manifold, there is a competition between
freely fluctuating configurations favored by entropy and
configurations bound to the defect, which are preferred by energy. This
competition can lead to a phase transition, the so called
delocalization or unbinding transition. It is often of second order,
that is, the amplitude of the fluctuations diverges continuously as
the transition point is approached from within the bound phase. This
leads to a scaling regime close to the transition whose universal
characteristics can be described by a continuum field theory.
Well-known examples of delocalization are wetting
phenomena~\cite{fish86}. For interfaces and polymers, these
transitions have been widely studied ~\cite{forg91}.  In the case of
polymers, even the generalized problem of $N$ mutually attracting
objects can be treated.  Analytically continued to
$N=0$, this describes a directed polymer in a random
medium~\cite{kard85}, which in turn is related to theories of
stochastic surface growth~\cite{kard86}.  The delocalization
transition then corresponds to a roughening transition between a
smooth and a rough growth mode.

An important class of low-dimensional manifolds are stretched objects
with mainly transversal fluctuations. These are described by a
$\dperp$-dimensional displacement field $\posvar(t)$ which depends on
a $\dpar$-dimensional internal variable $t$. The continuum
Hamiltonian takes the form
\begin{equation}
\label{H}
{\cal H}=
\int \left [\frac{1}{2} (\nabla^k \posvar)^2
+ V(\posvar, \nabla \posvar) \right ]
\,{\mathrm d}^{\dpar} t \;,
\end{equation}
where
$(\nabla^k \posvar)^2 \equiv
\sum_{\alpha = 1}^{\dpar}
\sum_{i = 1}^{\dperp} (\partial^k r_i (t) / \partial t_\alpha^k)^2$
is the leading tension ($k = 1$) or curvature ($k = 2$) energy in a
small-gradient expansion. The potential $V$ describes the interaction
of the manifold with an external object or boundary at $\posvar = 0$,
or the mutual interaction between
two manifolds with relative displacement $\posvar (t)$. Physical realizations
of flexible manifolds ($k = 1$) are  directed
polymers and flux lines in a type~II
superconductor ($\dpar = 1, \dperp = 2$)~\cite{blatt93,bale94},
steps on a tilted crystal surface ($\dpar = 1, \dperp = 1$),
and domain walls in a ferromagnet ($\dpar = 2, \dperp = 1$).
Short-ranged interactions of definite sign can be represented
as $V(\posvar(t)) = g \Phi (t)$ in terms of the local {\em contact field}
\begin{equation}
\Phi (t) \equiv \delta (\posvar(t)) \;.
\label{Phi}
\end{equation}
The scaling dimension of this field,
$x_\Phi = d \chi$,
is given in terms of the roughness exponent $\chi$, with $\chi = (2 - \dpar)/2$
for $k = 1$. There exists a well known perturbative
framework~\cite{dupl89,davi93} to treat such interactions, which has been
applied extensively to the problem of self-avoiding
manifolds~\cite{kard87,hwa90,dupl90,davi94,davi96}. More
complicated short-ranged~\cite{laes93} and long-ranged~\cite{laes96}
interactions have been studied as well.

In this Letter, we develop the field theory of semi-flexible
manifolds with local interactions, described by the Hamiltonian
(\ref{H}) with $k = 2$.
Physically interesting cases are again polymers
($\dpar = 1, \dperp = 1,2$)~\cite{magg89,lipo89a,gomp89} and,
in particular, fluid membranes ($\dpar = 2, \dperp = 1$)~\cite{lipo95}.
Since a semi-flexible manifold has a locally well-defined orientation,
we have to consider interactions $V (\posvar (t), \nabla \posvar (t))$
that depend both on the displacement and on the orientation.
We find there are now two important scaling fields:
Local contacts at arbitrary orientation are still represented by the
field $\Phi (t)$ given by (\ref{Phi}), which has dimension
$x_\Phi = d \chi$ with $\chi = (4 - \dpar)/2$. Due to the stiffness,
however, closeby contacts always have a preferred orientation parallel
to the defect. Such contacts are described by the field
\begin{equation}
\phinn (t) \equiv \delta (\posvar (t)) \delta (\nabla \posvar (t))
\label{Omega}
\end{equation}
\begin{figure}[t]
\narrowtext
\begin{center}
\epsfig{figure=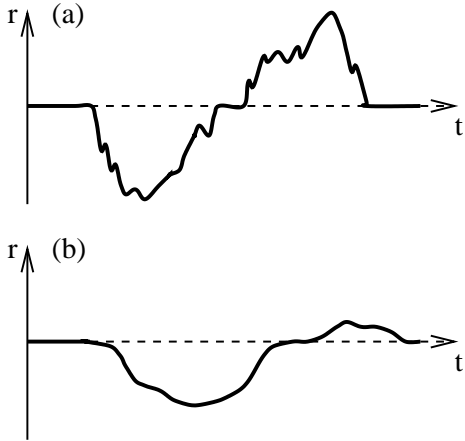,angle=0,width=0.7\columnwidth}
\caption{(a) A flexible and (b) a semiflexible polymer $r(t)$ bound
to an attractive defect indicated by the dashed lines. The unbound segments
join the  defect at arbitrary orientation and at fixed orientation
$ {\rm d} r / {\rm d}t = 0$, respectively.}
\end{center}
\end{figure}
\noindent of dimension
$x_\phinn = d\chi+d D(\chi-1)$. The scaling fields $\Phi$ and $\phinn$
are found to obey an operator product expansion.
This is a well-known concept in field theory (see, e.g.,
Ref.~\cite{card96}), which has been applied extensively to
flexible manifolds~\cite{lass98}.
It allows us to write down
renormalization group equations for generic local interactions.
This leads  to  results for the
delocalization of semi-flexible polymers and of fluid membranes.
Most importantly,
the bound state of a semi-flexible manifold turns out to be maintained by
contact interactions (\ref{Omega}) at fixed orientation, while the bound
state of a flexible manifold involves interactions of the form (\ref{Phi}).
Typical bound state configurations are compared in Fig.~1 for the case
of polymers ($D = 1$). Our one-loop results are in agreement
with previous results obtained by approximate
renormalization methods~\cite{lipo88,lipo89}
and by approaches specific to polymers~\cite{magg89,lipo89a,gomp89}.
They are exact for $D = 1$ and can be improved systematically for
higher values of $D$.
Furthermore, they can be applied to a semi-flexible manifold in a
quenched random potential (taken to be Gauss distributed
with mean
$\overline{ V(\posvar,t) } = 0$  and variance
$\overline{ V(\posvar,t)  V(\posvar',t')} =
 \sigma^2 \delta (t - t') \delta (\posvar - \posvar')$).
In the replica formalism, this is equivalent to
$N$ interacting semi-flexible manifolds in the limit of vanishing
$N$. For fluid membranes, any amount of
disorder is relevant and leads to a strong coupling phase. For
semi-flexible polymers, however, small amounts of disorder are
irrelevant (unlike for their flexible counterparts). There is now a
first-order transition to the strong coupling phase at a finite amount of
disorder. A quantitative
description of the disordered strong coupling phase is, however, beyond the
means of perturbation theory.

We study the manifold displacement field in a hypercube of longitudinal
extension
$0 \leq t_\alpha \leq T$ ($ \alpha = 1, \dots, \dpar$)
and of transversal extension
$0 \leq r_i \leq R$ ($ i = 1, \dots, \dperp$).
For $T\gg R^{1/\chi}$, the free energy becomes extensive,
$F \sim T^\dpar R^{-\dpar / \chi}$, and the perturbation series
for the free energy density $f \equiv F/T^D$ becomes invariant under
translations of $t$. This series takes the easiest form if the manifold
is subject to  {\em wall constraints} forcing the probability
density
$\rho(\posvar') \equiv
   \langle  \delta (\posvar (t) - \posvar') \rangle$
to vanish on the  boundary of the hypercube, in particular along the
``edge'' $\posvar = 0$.
The density then takes the asymptotic scaling form
$\rho(\posvar) \sim (r_1 \dots r_d)^\theta R^{- \dperp(1 + \theta)}$
for $|\posvar|\ll R$, with an
exponent $\theta > 0$ expressing long-ranged suppression of the
configurations close to the boundary.  The wall constraint is natural in
$\dperp=1$ for a fluid membrane at a planar system boundary, or for a pair
of membranes without mutual intersections. The generalization to
arbitrary $\dperp$ has been chosen such that the Hamiltonian~(\ref{H})
remains factorizable. Short-ranged interactions with the manifold
$\posvar = 0$ are now described by the local field
\begin{equation}
\phinn (t) \equiv \lim_{r \to 0} r^{- \dperp \theta}
                  \prod_{i = 1}^d \delta (r_i (t) - r) \;,
\label{Omega2}
\end{equation}
whose expectation values are finite. Due to the constraint, these
interactions are always at fixed orientation $\nabla \posvar = 0$. Hence, we
have used the same symbol $\Omega$ as for the field (\ref{Omega}) of
the unconstrained system. The correlation functions
\begin{eqnarray}
\langle \phinn (t) \rangle & \sim & R^{-x_{\phinn}/ \chi}
\label{Omegadens} \\
\langle \phinn (t) \phinn (t') \rangle & \sim &
  |t - t'|^{-x_{\phinn}} \langle \phinn (t) \rangle  + \dots
  \;\; (|t - t'|^\chi \ll R)
  \label{Omegapair}
\end{eqnarray}
define the scaling dimension $x_\Omega$. In the constrained system,
the  density (\ref{Omegadens}) is linked to the pressure of the system
by a wall theorem,
\begin{equation}
\langle \phinn
 \rangle  \sim \left(\partial f / \partial R\right)^\dperp
                            \sim R^{-\dperp (1 +  \dpar / \chi)} \;.
\end{equation}
This determines the exponent
value~\cite{Gompper}
\begin{equation}
x_{ \phinn} = d(\chi + D) \;,
\label{xOmega2}
\end{equation}
which differs from that of the unconstrained system.
It is in agreement with a conjecture
from functional renormalization~\cite{lipo95a} for general $\dpar$ and
a direct calculation~\cite{burk93} for $\dpar = 1$.

The interaction part $\delta f (h, R) \equiv f(h,R) - f(0,R)$
of the free energy of the system with
$V(\posvar(t),\nabla\posvar(t))=h\phinn(t)$
can be expanded as a power series
\begin{equation}
\delta f = h \langle \phinn \rangle
           +\frac{h^2}{2}\int\! {\rm d}^{\dpar}t
            \langle \phinn(0) \phinn(t) \rangle_c
           +O(h^3)
\label{fpert}\label{eq_freenser}
\end{equation}
containing the connected correlation functions
\begin{equation}
\langle \phinn(0) \phinn(t) \rangle_c\equiv
\langle \phinn(0) \phinn(t) \rangle-
\langle\phinn\rangle^2
\label{OOc}
\end{equation}
etc., taken at $h = 0$.  By (\ref{xOmega2}), there is a whole {\em line} in
the ($\dpar$,$\dperp$) plane,
where the interaction $\phinn$ is marginal, i.e., where
$x_{\phinn} =\dpar$. This line is given by
\begin{equation}\label{eq_marginalline}
\dperp^*(\dpar)=\frac{2\dpar}{4+\dpar}.
\end{equation}
The perturbation series~(\ref{eq_freenser}) has poles in
\begin{equation}
\epsilon\equiv \dpar - x_{\phinn} =\dpar-\dperp(2+\dpar/2) \;,
\end{equation}
which can be regularized around {\em any point} on the line of
marginality~(\ref{eq_marginalline}); see the discussion
in~\cite{kard87} for flexible manifolds. The singularity
of the two-point function (\ref{OOc}) is given by the
operator product expansion
\begin{equation}
\phinn(t) \phinn(t')   \sim  |t - t'|^{-x_\Omega} \phinn (t) +\dots.
\end{equation}
This singularity determines in a standard way~\cite{lass98} the
one-loop renormalization group equation of the
dimensionless coupling constant
${v} \equiv h R^{\epsilon/\chi}$. In an appropriate scheme, this takes the
form
\begin{equation}
\dot{{v}} = \epsilon\, {v} - {v}^2 + O(v^3) \;.
\label{beta}
\end{equation}
The unstable fixed point ${v}^* = \epsilon + O(\epsilon^2)$ represents the
transition. The linearized form
$ \dot{{v}} = \epsilon^* ({v} - {v}^*) + \dots$
with
$\epsilon^* = - \epsilon + O(\epsilon^2)$
then determines the scaling of
the transversal localization length
$\xi \equiv \langle \posvar^2 \rangle^{1/2}$,
\begin{equation}
\xi \sim ({v}^* - {v})^{-\chi/ \epsilon^*}
\qquad({v} < {v}^*) \;,
\label{xi}
\end{equation}
and the scaling dimension
\begin{equation}
x_{\phinn}^* = \dpar - \epsilon^* = 2 \dpar -
x_{\phinn} +O(\epsilon^2) \;,
\label{xO*}
\end{equation}
which takes the place of $x_{\phinn}$ in the correlations
(\ref{Omegadens}) and (\ref{Omegapair}) at the transition
point. These relations describe the scaling of a bound state maintained by
contact forces at fixed orientation. Typical configurations
look similar to those of Fig.~1(b) but are confined to the
region $r_i > 0$.

The most interesting application of
(\ref{xi}) and (\ref{xO*}) is the delocalization transition of a fluid
membrane from a hard wall ($\dpar = 2, \dperp = 1$), where $\xi \sim
(T_c - T)^{-1}$ (since the effective coupling is
temperature-dependent) and $x_{\phinn}^* = 1$. Not surprisingly,
these one-loop results are in agreement with those from
functional renormalization~\cite{lipo88}. They also fit very well the
numerical values of~\cite{lipo89}, which implies that higher order corrections
must be small. The system with wall constraint at $h = 0$ can be regarded as an
unconstrained system in the limit of large repulsive interaction. Conversely,
the scaling at the transition point of the constrained system may be related
to that of the free unconstrained system. Indeed, the one-loop value
$x^*_\Omega$ from (\ref{xO*}) equals the dimension $x_\Omega = 1$
of the free field (\ref{Omega}), indicating that the sum of the higher order
corrections in (\ref{beta}) and (\ref{xO*}) may vanish altogether
at the specific point $(D = 2, d = 1)$.

For the case of polymers ($D = 1$) it is easy to show that the multi-point
correlations entering (\ref{fpert})
factorize after ``time-ordering'' the interaction points,
\begin{equation}
\langle \phinn(t_1) \dots \phinn(t_n) \rangle =
  \langle \phinn(t_1)\rangle
  {\cal R} (t_2\! - t_1) \dots {\cal R} (t_n\! - t_{n-1})
\label{eq_factorize}
\end{equation}
for $t_1 < \dots < t_n$. The factors
${\cal R} (t) \equiv \langle \phinn(0)\phinn(t) \rangle
                     / \langle \phinn \rangle$
can be interpreted as ``return'' probabilities. Using (\ref{eq_factorize}),
one can show that the polymer perturbation series (\ref{fpert}) is
{\em one-loop renormalizable}; i.e., the connected two-point function
$\langle \phinn(0) \phinn(t) \rangle_c$
generates the only primitive singularity, and
there are no higher-order terms in (\ref{beta}). The arguments are completely
analogous to those for the case
$k = 1$~\cite{dupl89,lass98}. The implications of
(\ref{xi}) and (\ref{xO*}) on general semi-flexible polymers
are discussed and verified numerically in~\cite{bund99}.  Here we have
given a {\em unified derivation} of these relations, stressing the
theoretical analogies with their known counterparts for  $k =
1$~\cite{lipo91,lass98}.

\begin{figure}[ht]
\narrowtext
\begin{center}
\epsfig{figure=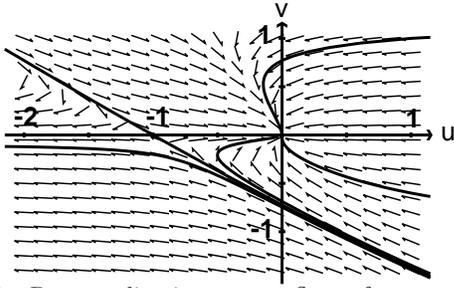,width=0.7\columnwidth}
\caption{Renormalization group flow of contact interactions
for a semi-flexible polymer in
$1+1$ dimensions. The flow equations (\ref{eq_uflow}), (\ref{eq_vflow})
have the single unstable
fixed point  $(u,v)=(-1,0)$ marking the delocalization transition.
\label{fig_flux}}
\end{center}
\end{figure}

We now turn to systems without the wall constraint and restrict ourselves to
$D=1$, namely mutually interacting semiflexible polymers  and, in particular,
a single such polymer in a random medium.
The latter system has a possible biostatistical application
in the theory of sequence alignment~\cite{hwa96}.
In the absence of a wall constraint,
we have to study generic contact interactions
$V(t) = g \Phi (t) + h \Omega (t)$ involving the fields
(\ref{Phi}) and~(\ref{Omega}). The perturbation series
 then contains connected correlations
$\langle \Phi(t_1) \dots \Phi (t_n) \phinn (t_1') \ldots\phinn (t_m') \rangle$
in the free theory ($g = h = 0$), which can be computed exactly.
They can be shown to obey the operator product expansion
\begin{eqnarray}
\Phi(t) \Phi(t') & \sim & |t - t'|^{-\dperp} \phinn (t) +\dots
\\
\Phi(t ) \phinn(t') & \sim & |t - t'|^{-\frac32\dperp} \phinn (t) +\dots
\\
\phinn(t) \phinn(t')  & \sim & |t - t'|^{-2\dperp} \phinn (t) +\dots.
\end{eqnarray}
These relations just say that {\em any} pair of closeby contacts amounts
to a single contact at fixed orientation, multiplied by a singular prefactor.
These singularities determine the one-loop
renormalization group equations
\begin{eqnarray}\label{eq_uflow}
\dot u & = & (1- 3d/2)u
\\
\dot v & = & \epsilon v - v^2 - u^2 - cuv
\label{eq_vflow}
\end{eqnarray}
for the dimensionless couplings $ u\equiv g R^{(1-3d/2)/\chi}$ and
$v\equiv hR^{\epsilon/\chi}$ with
\begin{equation}
\epsilon\equiv1-x_{\phinn}=1-2d.
\label{epsilon2}
\end{equation}
The corresponding flow diagram for $\dperp=1$ is shown in
Fig.~\ref{fig_flux}.  The unique delocalization fixed point $(u^* = 0,
v^* = \epsilon)$ is on the line $u = 0$. This property ensures
that the constant $c$ in (\ref{eq_vflow}) drops out of the critical
exponents. It will be preserved for any $d>2/3$ also
at higher orders since any operator product $\prod_{i,j} \Phi
(t_i) \phinn (t_{j}')$ couples only to $\phinn(t_1)$. The perturbation
series at $u = 0$, however, is factorizable according to
(\ref{eq_factorize}) and one-loop renormalizable in exactly the same
way as with the wall constraint.  Hence, the (in $D=1$) exact
relations~(\ref{xi}) and
(\ref{xO*}) still hold (with $\epsilon$ given by (\ref{epsilon2})
and $x_\Omega = 2d$),
resulting in $\xi \sim (T_{\mathrm{c}}-T)^{3/(2-4d)}$ for $2/3<d<1$
and $x_\phinn^* = 2-2d$.  This scaling dimension turning negative for
$\dperp > 1$ indicates that the transition becomes of first order; see
the discussion and extensive numerics in~\cite{bund99}. An analogous
first-order regime is known for flexible polymers~\cite{lipo91}.

The above arguments can be generalized to the replica theory
of $N$ semiflexible polymers coupled by pair contact forces.
If these are of type $\phinn$, the time-ordered perturbation series
can be mapped term by term onto the perturbation series of flexible
polymers with $\Phi$ interactions. Its leading divergencies
 are due to ``ladder'' diagrams with the same
pair of polymers interacting at subsequent points
$t_i$~\cite{lassig95,bund96}.
The  presence of $\Phi$ interactions
for semiflexible polymers does not change these singularities
by the same argument as for $N = 2$:
 Due to their stiffness, any two semi-flexible polymers
interacting twice in a short interval have to be parallel to each
other, or, in other words, the leading divergent diagrams
behave like diagrams involving only $\phinn$ operators.
For the $\Omega$ system, however, the results
of~\cite{lassig95,bund96} immediately carry over and imply that the
critical behavior at the delocalization transition  does not depend on $N$.
In particular, the random limit of vanishing $N$ becomes trivial. We conclude
that a $1+\dperp$ dimensional semi-flexible polymer in a random
potential has a phase
transition between a weak and a strong coupling phase at a critical
strength of the randomness for $\dperp>2/3$. This phase transition
corresponds to the roughening transition of the Kardar-Parisi-Zhang
equation in $4\dperp$ dimensions. For fluid membranes, on the other
hand, an arbitrarily small amount of disorder is relevant and leads to
a strong coupling phase.

We gratefully acknowledge useful discussions
with G. Gompper, C. Hiergeist, T. Hwa, and, in particular, with R. Lipowsky
and K. Wiese.

\mcend


\begin{references}
\vspace*{-0.5cm}

\bibitem{lipo89b}
For a review, see R. Lipowsky, Physica Scripta T{\bf 29}, 259 (1989).

\bibitem{fish86}
For a review, see M.E. Fisher, J. Chem. Soc. Faraday Trans. 2 {\bf 82},
1569 (1986).

\bibitem{forg91}
G. Forgacs, R. Lipowsky, and Th.M. Nieuwenhuizen, in {\it Phase Transitions
and Critical Phenomena}, Vol. 14, (Acadamic Press, London, 1991), and
references therein.

\bibitem{kard85}
M. Kardar, Phys. Rev. Lett. {\bf 55}, 2235 (1985); Nucl. Phys. B {\bf 290},
582 (1987).

\bibitem{kard86}
M. Kardar, G. Parisi, and Y.-C. Zhang, Phys. Rev. Lett.
{\bf 56}, 889 (1986).

\bibitem{blatt93}
G.Blatter, M.V.Feigel'man, V.B.Geshkenbein, A.I. Lar\-kin, and V.M. Vinokur,
Physica A {\bf 200}, 341 (1993).

\bibitem{bale94}
L. Balents and M. Kardar, Phys. Rev. B {\bf 49}, 13030 (1994).

\bibitem{dupl89}
B. Duplantier, Phys. Rev. Lett. {\bf 62}, 2337 (1989).

\bibitem{davi93}
F. David, B. Duplantier, and E. Guitter, Phys. Rev. Lett. {\bf 70}, 2205
(1993); Nucl. Phys. B {\bf 394}, 555 (1993).

\bibitem{kard87}
M. Kardar and D.R. Nelson, Phys. Rev. Lett. {\bf 58}, 1289 (1987).

\bibitem{hwa90}
T. Hwa, Phys. Rev. A {\bf 41}, 1751 (1990).

\bibitem{dupl90}
B. Duplantier, T. Hwa, and M. Kardar, Phys. Rev. Lett. {\bf 64}, 2022 (1990).

\bibitem{davi94}
F. David, B. Duplantier, and E. Guitter, Phys. Rev. Lett. {\bf 72}, 311 (1994).

\bibitem{davi96}
F. David and K.J. Wiese, Phys. Rev. Lett. {\bf 76}, 4564 (1996);
Nucl. Phys. B {\bf 487}, 529 (1997).

\bibitem{laes93}
M. L{\"a}ssig and R. Lipowsky, Phys. Rev. Lett. {\bf 70}, 1131 (1993).

\bibitem{laes96}
M. L{\"a}ssig, Phys. Rev. Lett. {\bf 77}, 526, 1996.

\bibitem{magg89}
A.C. Maggs, D.A. Huse, and S. Leibler, Europhys. Lett. {\bf 8}, 615 (1989).

\bibitem{lipo89a}
R. Lipowsky, Phys. Rev. Lett. {\bf 62}, 704 (1989).

\bibitem{gomp89}
G. Gompper and T.W. Burkhardt, Phys. Rev. A {\bf 40}, 6124 (1989).

\bibitem{lipo95}
R. Lipowsky, in: R. Lipowsky, E. Sackmann (eds.), {\it Structure and Dynamics
of Membranes, Handbook of Biological Physics} {\bf 1}, Elsevier,
Amsterdam, 521 (1995).

\bibitem{card96}
J. Cardy, {\it Scaling and Renormalization in Statistical Physics}
(Cambridge University Press, Cambridge, 1996).

\bibitem{lass98}
For a review, see M. L{\"a}ssig, J. Phys. C {\bf 10}, 9905 (1998).

\bibitem{lipo88}
R. Lipowsky, Europhys. Lett. {\bf 7}, 255 (1988).

\bibitem{lipo89}
R. Lipowsky and B. Zielinska, Phys. Rev. Lett. {\bf 62}, 1572 (1989).

\bibitem{Gompper}
G. Gompper, private communication;
G. Gompper and D. Kroll, J. Phys I France {\bf 1}, 1411 (1991).

\bibitem{lipo95a}
R. Lipowsky, Z. Phys. B {\bf 97}, 193 (1995).

\bibitem{burk93}
T.W. Burkhardt, J. Phys. {\bf A 26}, L1157 (1993).

\bibitem{bund99}
R. Bundschuh, M. L\"assig, and R. Lipowsky, preprint (1999).

\bibitem{lipo91}
R. Lipowsky, Europhys. Lett. {\bf 15}, 703 (1991).

\bibitem{hwa96}
T. Hwa and M. L{\"a}ssig, Phys. Rev. Lett. {\bf 76}, 2591 (1996).

\bibitem{lassig95}
M. L\"assig, Nucl. Phys. B {\bf 448}, 559 (1995).

\bibitem{bund96}
R. Bundschuh and M. L{\"a}ssig, Phys. Rev. E {\bf 54}, 304 (1996).

\end{references}
\end{document}